\documentclass[preprint2]{aastex63}
\usepackage{epsfig}
\usepackage{amsmath}

\shortauthors{J. Li et al.}
\shorttitle{Chandra Observations of NGC~4438}

\begin{document}

\title{CHANG-ES XXIX: The Sub-kpc Nuclear Bubble of NGC~4438}

\author[0000-0001-6239-3821]{Jiang-Tao Li}
\affiliation{Department of Astronomy, University of Michigan, 311 West Hall, 1085 S. University Ave, Ann Arbor, MI, 48109-1107, U.S.A.}
\affiliation{Purple Mountain Observatory, Chinese Academy of Sciences, 10 Yuanhua Road, Nanjing 210023, China}

\author{Q. Daniel Wang}
\affiliation{Department of Astronomy, University of Massachusetts, 710 North Pleasant Street, Amherst, MA, 01003, U.S.A.}

\author{Theresa Wiegert}
\affiliation{Instituto de Astrof$\acute{i}$sica de Andaluc$\acute{i}$a (IAA-CSIC), Glorieta de la Astronom$\acute{i}$a, 18008, Granada, Spain}

\author{Joel N. Bregman}
\affiliation{Department of Astronomy, University of Michigan, 311 West Hall, 1085 S. University Ave, Ann Arbor, MI, 48109-1107, U.S.A.}

\author{Rainer Beck} 
\affiliation{Max-Planck-Institut f$\ddot{u}$r Radioastronomie, Auf dem H$\ddot{u}$gel 69, 53121 Bonn, Germany}

\author{Ancor Damas-Segovia}
\affiliation{Instituto de Astrof$\acute{i}$sica de Andaluc$\acute{i}$a (IAA-CSIC), Glorieta de la Astronom$\acute{i}$a, 18008, Granada, Spain}

\author{Judith A. Irwin}
\affiliation{Department of Physics, Engineering Physics, \& Astronomy, Queens University, Kingston, ON, K7L 3N6, Canada}

\author{Li Ji}
\affiliation{Purple Mountain Observatory, Chinese Academy of Sciences, 10 Yuanhua Road, Nanjing 210023, China}

\author{Yelena Stein}
\affiliation{Ruhr-Universit$\ddot{a}$t Bochum, Fakult$\ddot{a}$t f$\ddot{u}$r Physik und Astronomie, Astronomisches Institut (AIRUB), Universit$\ddot{a}$tsstrasse 150, 44801 Bochum, Germany}

\author{Wei Sun}
\affiliation{Purple Mountain Observatory, Chinese Academy of Sciences, 10 Yuanhua Road, Nanjing 210023, China}

\author{Yang Yang}
\affiliation{Purple Mountain Observatory, Chinese Academy of Sciences, 10 Yuanhua Road, Nanjing 210023, China}
\affiliation{School of Astronomy and Space Science, Nanjing University, Nanjing 210023, People’s Republic of China}

\correspondingauthor{Jiang-Tao Li}
\email{pandataotao@gmail.com}

\begin{abstract}
AGN bubbles could play an important role in accelerating high-energy CRs and galactic feedback. Only in nearby galaxies could we have high enough angular resolution in multi-wavelengths to study the sub-kpc environment of the AGN, where the bubbles are produced and strongly interact with the surrounding ISM. In this paper, we present the latest \emph{Chandra} observations of the Virgo cluster galaxy NGC~4438, which hosts multi-scale bubbles detected in various bands. The galaxy also has low current star formation activity, so these bubbles are evidently produced by the AGN rather than a starburst. We present spatially resolved spectral analysis of the \emph{Chandra} data of the $\sim3^{\prime\prime}\times5^{\prime\prime}$ ($\sim200{\rm~pc}\times350\rm~pc$) nuclear bubble of NGC~4438. The power law tail in the X-ray spectra can be most naturally explained as synchrotron emission from high-energy CR leptons. The hot gas temperature increases, while the overall contribution of the non-thermal X-ray emission decreases with the vertical distance from the galactic plane. We calculate the synchrotron cooling timescale of the CR leptons responsible for the non-thermal hard X-ray emission to be only a few tens to a few hundreds of years. The thermal pressure of the hot gas is about three times the magnetic pressure, but the current data cannot rule out the possibility that they are still in pressure balance. The spatially resolved spectroscopy presented in this paper may have important constraints on how the AGN accelerates CRs and drives outflows. We also discover a transient X-ray source only $\sim5^{\prime\prime}$ from the nucleus of NGC~4438. The source was not detected in 2002 and 2008, but became quite X-ray bright in March 2020, with an average 0.5-7~keV luminosity of $\sim10^{39}\rm~ergs~s^{-1}$. 
\end{abstract}

\keywords{galaxies: active - Galaxies, galaxies: jets - Galaxies, ISM: bubbles - Interstellar Medium (ISM), Nebulae, (ISM:) cosmic rays - Interstellar Medium (ISM), Nebulae, X-rays: galaxies - Resolved and unresolved sources as a function of wavelength, ISM: jets and outflows - Interstellar Medium (ISM), Nebulae}


\section{Introduction} \label{sec:Intro}


Bubbles in the nuclear regions of galaxies are important signatures of galactic feedback from either starbursts or AGN (e.g., \citealt{Su10,Guo12,Crocker15}). Unlike the starburst-driven superbubbles that are typically $\sim\rm kpc$ or tens of kpc scale (e.g., \citealt{Li08,Li19}), the AGN-driven bubbles could be largely diverse in physical scale and properties. Some AGN-driven bubbles or cavities could be a few tens or even a few hundreds of kpc in size --- energetic enough to affect the gaseous medium of the entire galaxy cluster (e.g., \citealt{Fabian00,Zhuravleva14,Gill21}). On the other hand, there are also AGN bubbles with a physical size more than three orders of magnitude smaller (e.g., \citealt{Machacek04,Irwin17}), which could strongly impact the interstellar medium (ISM) in the galactic nuclear region, including not only the multi-phase gases, but also the dust, cosmic-rays (CRs), and the magnetic field.

AGN bubbles are potential factories of high energy CRs, especially those above the ``knee'' of the CR energy spectrum (at $\sim10^{15}\rm~eV$) which cannot be accelerated within individual supernova remnants (e.g., \citealt{Kotera11} and references therein). These high energy CRs could be important in driving AGN outflows, which plays an important role in regulating the co-evolution of galaxies, their environments, and the large scale structure (e.g., \citealt{Uhlig12,Ruszkowski17,Farber18,Holguin19}). Most of the large scale features related to AGN bubbles are cavities of X-ray emitting hot gas (e.g., \citealt{Fabian00,Zhuravleva14}) or radio structures produced by the emissions of low energy CRs (e.g., \citealt{Gill21}). These features have been largely reshaped by the environment, so they are not direct probes of the central engine. We need observations of small scale bubbles directly related to the AGN, in order to better understand how AGN feedback works, such as the acceleration, propagation, and radiative cooling of CRs, and the energy equipartition between different ISM phases. Although extremely high-energy CRs or photons have been detected from AGN, it is typically difficult to spatially resolve the fine structures in these high energy observations (e.g., \citealt{Madejski16} and references therein). On the other hand, it is much easier to resolve galactic nuclear bubbles at longer wavelengths, such as soft X-ray, UV/optical/near-IR, and radio bands (e.g., \citealt{Kenney02,Machacek04,Irwin17}). Some of these bands are dominated by thermal emissions from different gas phases, while most of the non-thermal emissions from CRs are detected in the radio band, which are produced by low-energy CR leptons (typically in GeV energy range, depending on the band) with a long radiative cooling timescale (typical value $t_{\rm cool}\sim10^{8-10}\rm~yr$; e.g., \citealt{Krause18}). 

Among all the electromagnetic wave bands, \emph{Chandra} observations in the hard X-ray band ($\sim2-7\rm~keV$) are probably the best for studying high energy CRs produced in galactic nuclear bubbles. It has arcsec angular resolution to resolve the small-scale structures, and traces non-thermal emission produced by $\sim(10-100)\rm~TeV$ CRs with short radiative cooling timescale (typical value $t_{\rm cool}\sim10^{1-3}\rm~yr$; e.g., \citealt{Li19}). The best example of spatially resolved hard X-ray observation of a galactic nuclear bubble is NGC~3079, where extended diffuse hard X-ray emission from the southern half of the kpc-scale bubbles has been detected (\citealt{Li19}; also see detections of diffuse hard X-ray emission in other galaxies in \citealt{Lacki13}, where the synchrotron origin of the emission is not as explicit). The synchrotron cooling timescale of the CR leptons responsible for the diffuse hard X-ray emission is too short so the CRs must be accelerated in situ, rather than accelerated by the AGN and subsequently propagating to the bubble shell. This is the first direct evidence for CR acceleration by a galactic nuclear bubble in an external galaxy.


In this paper, we present spatially resolved hard X-ray observations of another galactic nuclear bubble, residing in the Virgo cluster galaxy NGC~4438 ($d=14.4\rm~Mpc$, $1^{\prime\prime}\approx70\rm~pc$). The soft X-ray emission from this galaxy has been well studied, and shows a lopsided $\sim2^\prime$ diameter bubble on the western side of the warped galactic disk \citep{Machacek04}. The galaxy also appears to be significantly brighter in X-ray than other galaxies with a comparable stellar feedback rate (from both old and young stellar populations; \citealt{Li13a,Li13b,Wang16}), which indicates that additional gas and/or heating sources are present. Compared to NGC~3079 studied in \citet{Li19}, NGC~4438 provides us with a unique laboratory to study galactic nuclear bubbles and their relation to feedback, because: (1) NGC~4438 hosts bubbles on clearly distinguishable scales: a $\sim3^{\prime\prime}\times5^{\prime\prime}$ nuclear bubble with a fainter counter bubble and a $d\sim2.5^\prime$ lopsided bubble only on the western side. Both bubbles are clearly detected at radio frequencies \citep{Hummel91,Hota07}, in X-ray \citep{Machacek04}, and by optical emission lines \citep{Kenney02,Kenney08} (see Fig.~\ref{fig:NGC4438bubblesImg} later in the present paper). (2) NGC~4438 lies in the M86 group, a subgroup of the Virgo cluster. The galaxy not only has tidal interactions with its close companion NGC~4435, as indicated by the warped disk and tidal tails, but is also connected to M86 via cold and hot gas filaments (e.g., \citealt{Kenney08}). The high density environment makes it ideal to examine the effect of feedback. (3) The galaxy shows little recent star formation \citep{Vargas19}, and the well resolved nuclear bubble strongly suggests it is likely that the bubble was produced by the AGN. This makes the explanation of the observations less confused by a nuclear starburst as is often the case in other galaxies (e.g., NGC~3079; \citealt{Li19}).

The present paper is organized as follows: We will present the \emph{Chandra} observations, data analysis, a brief description of the multi-wavelength data used for comparison, and X-ray analysis of some interesting nuclear point-like sources in \S\ref{Sec:ObsDataReduc}. We will then conduct spatially resolved X-ray spectroscopy analysis of the nuclear bubble and discuss the results in \S\ref{sec:NuclearBubbleN4438}. Our main results and conclusions will be summarized in \S\ref{Sec:SummaryConclusion}. Unless explicitly noted, all errors are quoted at 1~$\sigma$ confidence level throughout the paper.


\section{Observations and Data Analysis} \label{Sec:ObsDataReduc}

\subsection{Chandra Data Calibration} \label{subsec:ChandraDataCalibration}

The NGC~4438/M86 area has been observed by \emph{Chandra} many times. Since the major focus of the present paper is the nuclear bubble of NGC~4438, we only select \emph{Chandra} observations covering the nuclear region of NGC~4438. There are six \emph{Chandra} observations used in the present study, which are summarized in Table~\ref{table:ChandraObs}. The first one, taken in 2002 ($\sim25\rm~ks$; ObsID 2883), has been published in \citet{Machacek04} and some other later papers (e.g., \citealt{Li13a,Li13b}), while ObsID 8042 is a snapshot observation ($\sim5\rm~ks$). The main new data used in this paper are the four observations taken in March 2020, with a total exposure time of $\sim95\rm~ks$ (from a \emph{Chandra} Cycle~20 program; PI: Li). All these six observations were taken with the \emph{Chandra}/ACIS in imaging mode, with NGC~4438 located on the S3 chip.


\begin{table}
\begin{center}
\caption{\emph{Chandra} Observations of NGC~4438} 
\footnotesize
\tabcolsep=3.pt%
\begin{tabular}{lcccccccccccccc}
\hline\hline
ObsID & PI & Mode & Start Date & $t_{\rm exp}$/ks \\
\hline
2883 & Jones & VFAINT & 2002-01-29 17:22:45 & 25.07 \\
8042 & Treu & FAINT & 2008-02-11 19:56:06 & 4.90 \\
21376 & Li & FAINT & 2020-03-20 04:02:24 & 29.68 \\
23189 & Li & FAINT & 2020-03-20 23:34:00 & 19.79 \\
23037 & Li & FAINT & 2020-03-27 11:55:56 & 19.79 \\
23200 & Li & FAINT & 2020-03-28 01:35:54 & 25.71 \\
\hline\hline
\end{tabular}\label{table:ChandraObs}
\end{center}
\end{table}

The \emph{Chandra} data are reprocessed with {\small CIAO} v4.13, {\small CALDB} v4.9.5, and the latest ACIS background files. We reprocess the evt1 raw data with the {\small CIAO} tool {\small chandra\_repro}. We then reproject and combine different observations to create a merged event list and exposure-corrected images with the {\small CIAO} tool {\small merge\_obs}. The images are extracted in standard ACIS Science Energy Bands (broad: 0.5-7.0~keV; ultrasoft: 0.2-0.5~keV; soft: 0.5-1.2~keV; medium: 1.2-2.0~keV; hard: 2.0-7.0~keV) used in the \emph{Chandra} source catalog (e.g., \citealt{Evans10}). 


Near the optical axis, the core size of the \emph{Chandra} point spread function (PSF) is smaller than the ACIS pixel size ($\approx0.492^{\prime\prime}$). Therefore, we adopt the ACIS Energy-Dependent Subpixel Event Repositioning (EDSER) algorithm \citep{Li04} to recalibrate the evt1 data, in order to enable subpixel resolution of some arcsec-scale features (e.g., the $\approx3^{\prime\prime}\times5^{\prime\prime}$ nuclear bubble of NGC~4438). The EDSER algorithm is applied by rerunning {\small chandra\_repro} with the setting ``pix\_adj=edser''. The images are then rebinned to a resolution of 0.25~pixel ($\approx0.123^{\prime\prime}$). Adopting the EDSER algorithm has little effect on sources far from the optical axis or any features without small-scale structures. Since the \emph{Chandra} PSF has not been calibrated on subpixel scales, we only adopt this algorithm when it is critical to improve the spatial analysis, i.e., in the spatially resolved spectroscopy analysis of the nuclear bubble of NGC~4438 (\S\ref{sec:NuclearBubbleN4438}). Examples of the \emph{Chandra} images created without applying the EDSER algorithm are presented in Figs.~\ref{fig:NGC4438src01}a-f and \ref{fig:NGC4435nuclear}a, while the subpixel images created with this algorithm are presented in Fig.~\ref{fig:NGC4438bubblesImg}e,f for comparison.


\begin{table*}
\begin{center}
\rotatebox{90}{
\begin{minipage}{\textheight}
\caption{Spectral Analysis Results of Different Objects} 
\scriptsize
\tabcolsep=3.pt%
\begin{tabular}{lcccccccccccccc}
\hline\hline
Object & Model & $N_{\rm H}$ & $kT$ & $EM_{\rm hot}$ & $n_{\rm e}$ & $P_{\rm hot}$ & $\Gamma$ & $\chi^2/\rm d.o.f.$ \\
 &  & $10^{20}\rm~cm^{-2}$ & keV & $\rm 10^{-3}cm^{-6}kpc^3$ & $f^{-1/2}\rm cm^{-3}$ & $10^{-13}f^{-1/2}\rm eV~cm^{-3}$&  \\
\hline
NGC~4438 variable source & power & 2.13 (fixed) & - & - & - & - & $2.36\pm0.12$ & 50.90/37 \\
\hfill (J122745.4+130028.5)\hfill &  &  & & & &  &  &  \\
NGC~4435 src01 & APEC+power & $96_{-50}^{+109}$ (thermal)$^{\rm I}$ & $0.07_{-0.03}^{+0.09}$ & - & - & - & $1.51_{-0.39}^{+3.36}$ & 8.28/10 \\
\hfill (J122740.5+130444.0, AGN)\hfill & & $283_{-171}^{+204}$ (power) &  & & & &  &  \\
NGC~4435 src02 & power & 2.13 (fixed) & - & - & - & - &  $1.61\pm0.13$ & 24.77/21 \\
\hfill (J122740.8+130447.5)\hfill &  &  & & & &  &  &  \\
NGC~4438 nuclear bubble & APEC+power & $14\pm3$ & $0.90\pm0.03$ & $4.2_{-0.3}^{+0.4}$ & $0.79\pm0.03$ & $617\pm32$ & $2.74_{-0.18}^{+0.19}$ & 163.12/156 \\
 & 2T (low-T) & $49\pm5$ & $0.27_{-0.01}^{+0.02}$ & $35_{-10}^{+14}$ & $2.3_{-0.3}^{+0.5}$ & $531_{-81}^{+112}$ & - & 193.77/156 \\
 & 2T (high-T) & - & $1.22_{-0.04}^{+0.05}$ & $7.9_{-0.5}^{+0.3}$ & $0.79\pm0.02$ & $826_{-37}^{+39}$ & - & - \\
NGC~4438 bubble A & APEC+power & $70_{-24}^{+11}$ & $0.61_{-0.13}^{+0.10}$ & $4.2_{-0.3}^{+0.2}$ & $1.29_{-0.44}^{+0.29}$ & $680_{-273}^{+189}$ & $1.38_{-1.38}^{+1.34}$ & 24.32/20 \\
NGC~4438 bubble B & APEC+power & $30_{-9}^{+8}$ & $0.83_{-0.05}^{+0.09}$ & $2.1_{-0.6}^{+0.8}$ & $0.91_{-0.12}^{+0.17}$ & $656_{-95}^{+140}$ & $2.58_{-0.42}^{+0.39}$ & 56.13/48 \\
NGC~4438 bubble C & APEC+power & $8~(<26)$ & $0.96_{-0.06}^{+0.05}$ & $1.2_{-0.2}^{+0.7}$ & $0.69_{-0.06}^{+0.21}$ & $573_{-61}^{+177}$ & $2.92_{-1.13}^{2.08}$ & 34.28/35 \\
NGC~4438 bubble D & APEC+power & $52_{-5}^{+6}$ & $0.34_{-0.04}^{+0.03}$ & $12_{-3}^{+6}$$^{\rm II}$ & $1.57_{-0.19}^{+0.37}$ & $464_{-74}^{+120}$ & $1.09~(<5)^{\rm III}$ & 27.71/22 \\
\hline\hline\\
\end{tabular}\label{table:SpecPara}
For point-like sources in the nuclear regions of NGC~4435 and NGC~4438, we also list their J names in the format of J$hhmmss.s\pm ddmmss.s$. All the spectra are fitted with a power law model with a photon index $\Gamma$. Whenever applicable, there is also an APEC component representing the thermal emission, with the hot gas temperature $kT$ and volume emission measure $EM_{\rm hot}$. The volume emission measure is defined as $EM_{\rm hot}\equiv\int n_{\rm e}n_{\rm H}fdV$, where $f$ is the volume filling factor and is assumed to be constant within the emitting volume. $EM_{\rm hot}$ is directly linked to the normalization of the APEC model in the form of $norm=\frac{10^{-14}}{4\pi[D_{\rm A}(1+z)]^2}$, where $D_{\rm A}$ is the angular diameter distance to the source in cm, while $n_{\rm e}$ and $n_{\rm H}$ are the electron and hydrogen number densities in $\rm cm^{-3}$. The thermal pressure $P_{\rm hot}$ is derived from $kT$ and $n_{\rm e}$. We adopt $\rm eV~cm^{-3}$ as the unit of pressure throughout the paper, in order to provide the readers a direct sense of the energy density. The cgs pressure unit $\rm dyn~cm^{-2}$ is linked to $\rm eV~cm^{-3}$ in the form of $\rm 1~dyn~cm^{-2}=6.24150913\times10^{11}~eV~cm^{-3}$. $N_{\rm H}$ is the foreground absorption column density of the ``tbabs'' model applied to both the APEC and power law components, except for the AGN of NGC~4435 (J122740.5+130444.0), where $N_{\rm H}$ has different values for the two components. $f$ is the unknown volume filling factor used to calculate $n_{\rm e}$ and $P_{\rm hot}$. $f$ is $\leq1$ by definition. We test two different models for the entire nuclear bubble: an ``APEC+power law'' model shown in Fig.~\ref{fig:NGC4438bubbleModel}a and a ``2T'' (APEC+APEC) model shown in Fig.~\ref{fig:NGC4438bubbleModel}b. The key parameters of the two models are both presented in the table for comparison. We adopt the ``APEC+power law'' model in the analysis of the other regions of the nuclear bubble.\\
\\
I: The two components have different $N_{\rm H}$ values.\\
II: The high $EM_{\rm hot}$, as well as other derived parameters ($n_{\rm e}$), is apparently caused by the unusually low $kT$, which could be biased due to the $kT$-$N_{\rm H}$ degeneracy as shown in Fig.~\ref{fig:NGC4438bubbleParaSpace}d and discussed in \S\ref{sec:NuclearBubbleN4438}.\\
III: $\Gamma$ is poorly constrained when the overall contribution of the power law component is low, so we set a fixed upper limit.
\end{minipage}
}
\end{center}
\end{table*}

\subsection{X-ray Sources in the Nuclear Region of NGC~4435 and NGC~4438} \label{subsec:NuclearSources}

\begin{figure*}[!th]
\begin{center}
\epsfig{figure=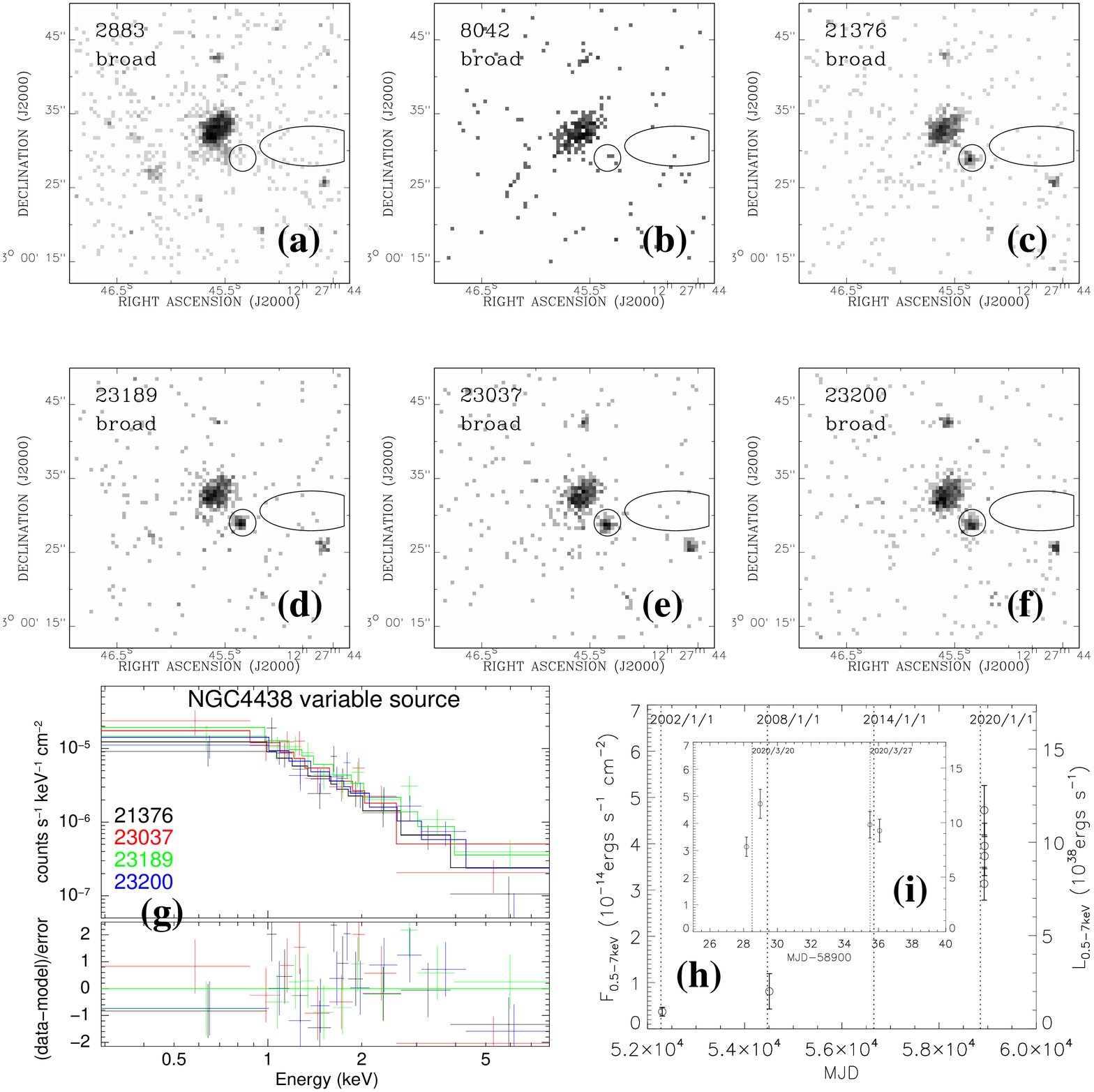,width=1.0\textwidth,angle=0, clip=}
\caption{Images (a-f), spectra (g), and lightcurve (h,i) of the variable source J122745.4+130028.5 in the nuclear region of NGC~4438. These images are constructed around the optical center of the galaxy ($0.6^\prime\times0.6^\prime$) in the 0.5-7~keV band and exposure-corrected (in unit of $\rm photons~s^{-1}~cm^{-2}$) for individual observations (their ObsIDs are denoted at top left in each panel). The source is marked with a circle of a radius $r=1.8^{\prime\prime}$, while the elliptical region (cropped at the edge of the FOV) is used to extract the local background spectra. Different colors in panel~(g) denote spectra extracted from the four different observations taken in March 2020 and jointly fitted with a power law model subjected to Galactic foreground absorption (no additional absorption was found). Panel~(h) is the light curve of all the six observations covering the source, with the left and right axes denoting the 0.5-7~keV flux and luminosity (assuming a distance of NGC~4438) of the source, respectively. Panel~(i) is a zoom-in of (h), covering only the four observations taken in March 2020.
}\label{fig:NGC4438src01}
\end{center}
\end{figure*}

We first look at the most prominent point-like X-ray sources in the nuclear regions of NGC~4435 and NGC~4438, while X-ray point sources in the surrounding area will be presented in a future paper. As shown in Fig.~\ref{fig:NGC4438src01}a-f and Fig.~\ref{fig:NGC4438bubblesImg}e,f, although we do not detect any significant X-ray point sources apparently associated with the galactic nucleus of NGC~4438, there is an X-ray bright bubble extending along the southeast (SE)-northwest (NW) direction in the nuclear region, with a semi-minor/semi-major diameter of $\sim3^{\prime\prime}/5^{\prime\prime}$. Spatially resolved X-ray emission from this bubble has been first reported in \citet{Machacek04}, and will be further discussed in \S\ref{sec:NuclearBubbleN4438}. In addition to this nuclear bubble and its counterpart in the opposite direction (the ``counter bubble''; Fig.~\ref{fig:NGC4438bubblesImg}c), there are also a few bright point-like sources distributed in the nuclear region of NGC~4438, which are possibly associated with the galaxy. In particular, with the new observations taken in 2020, we discovered a transient source located at J122745.4+130028.5, which did not show any significant X-ray emission in 2002 and 2008, but appear quite significant in all observations taken in 2020.

We extract the \emph{Chandra} spectra of J122745.4+130028.5 from a $r=1.8^{\prime\prime}$ circular region in each observations, using the {\small CIAO} tool {\small specextract}. The corresponding sky background is extracted from an adjacent elliptical region apparently free of diffuse X-ray emissions (Fig.~\ref{fig:NGC4438src01}a-f). We choose a small local background region, instead of a larger sky background region located far away from the galaxy (such as adopted in the analysis of the nuclear bubble below). This is because the source appears very faint in some epochs, and the diffuse emission from the nuclear bubble may seriously affect the measurement of the source flux. Only the four observations taken in 2020 have sufficient numbers of photons from this variable source. We then jointly fit these four \emph{Chandra} spectra with an absorbed power law model (Fig.~\ref{fig:NGC4438src01}g). We do not find significant additional foreground absorption from the galaxy, so the absorption column density has been fixed at the Milky Way foreground value of $N_{\rm H}=2.13\times10^{20}\rm~cm^{-2}$, which is obtained from the HEASARC web tools (data from \citealt{Dickey90,Kalberla05,BenBekhti16}). As the power law photon index does not show significant variation among these four observations, they are linked to an identical value. The only free parameters are thus a single linked power law photon index, and the X-ray flux in four different epochs. The best-fit photon index is $\Gamma=2.36\pm0.12$ (Table~\ref{table:SpecPara}). We also compute the X-ray flux at the six epochs to confirm the variability of the source (Fig.~\ref{fig:NGC4438src01}h). As there are insufficient numbers of photons to extract a spectrum from the first two observations, we simply convert the observed net counts rate to the X-ray flux, assuming the same absorbed power law model as the later four observations. To be consistent, the X-ray fluxes of the later four observations are also computed this way, and are consistent with those obtained from the spectral analysis. We also compute the X-ray luminosity at each epoch, assuming the source is at the same distance as NGC~4438 ($d=14.4\rm~Mpc$). As shown in Fig.~\ref{fig:NGC4438src01}h, the X-ray flux of the first two observations is still above zero, which may be caused by the excess X-ray emission in the nuclear region as compared to the background region, but it is still clear that the X-ray emission from the source was much stronger in 2020. The average 0.5-7~keV luminosity of J122745.4+130028.5 in March 2020 is $\sim10^{39}\rm~ergs~s^{-1}$, assuming it is located inside the galaxy. The X-ray flux does not show significant variation among the four observations taken within about one week (Fig.~\ref{fig:NGC4438src01}i).

\begin{figure*}[!th]
\begin{center}
\epsfig{figure=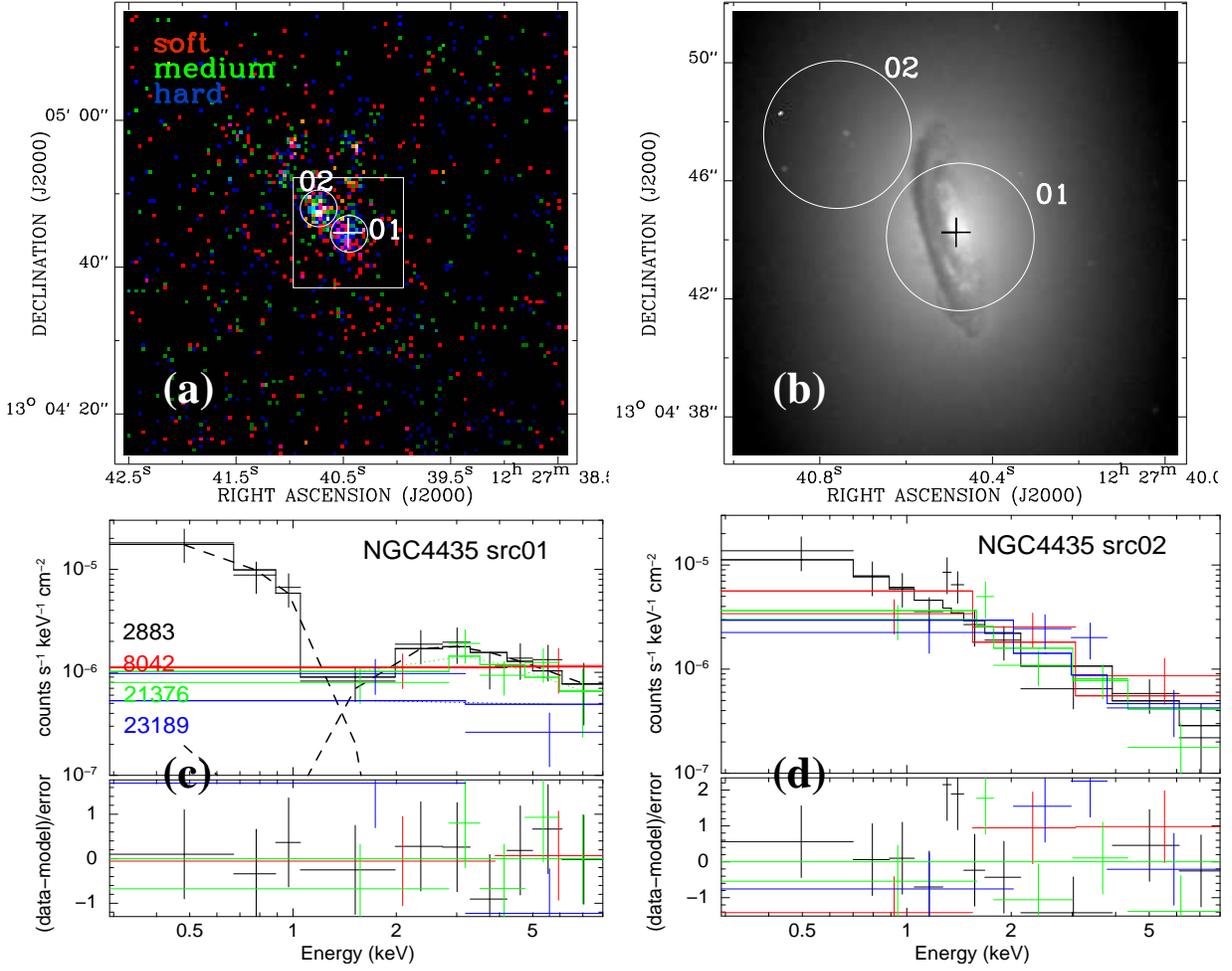,width=0.9\textwidth,angle=0, clip=}
\caption{(a) \emph{Chandra} tri-color images of the central $1^\prime\times1^\prime$ of NGC~4435, the companion galaxy at NW of NGC~4438. The white plus marks the optical center of the galaxy. The two circles mark the two brightest X-ray sources in the nuclear region. The $15^{\prime\prime}\times15^{\prime\prime}$ box is the FOV of the \emph{HST}/F475W image shown in panel~(b), where the same circles are plotted to show the positions of the two X-ray sources. It is clear that src01 coincides with the nucleus of the galaxy. Panels (c) and (d) show the \emph{Chandra} spectra of the two sources fitted with an absorbed power law model; src01 (c) also has a thermal component; individual model components are plotted as dashed curves. The nuclear region of NGC~4435 is only covered by four observations, which are denoted in the corresponding colors in (c). 
}\label{fig:NGC4435nuclear}
\end{center}
\end{figure*}

Combining all the \emph{Chandra} observations, the X-ray emission in the nuclear region of NGC~4435 is resolved into a few point-like sources, with some possible excess in diffuse emission (Fig.~\ref{fig:NGC4435nuclear}a). The two brightest sources, J122740.4+130444.0 (NGC~4435 src01) and J122740.7+130447.5 (NGC~4435 src02), are separated by only $\approx5.4^{\prime\prime}$, with src01 coinciding with the optical nucleus surrounded by a small dusty disk (with a diameter of $\approx7.6^{\prime\prime}$), which is well resolved on the \emph{HST} images (Fig.~\ref{fig:NGC4435nuclear}b). Src01 is thus likely the X-ray counterpart of the AGN. 

The nuclear region of NGC~4435 is not covered in two of the observations (ObsID=23037 and 23200). We therefore extract the \emph{Chandra} spectra from the other four observations, from two circular regions of $r=2.5^{\prime\prime}$ (the larger size than J122745.4+130028.5 in NGC~4438 is due to the larger off-axis PSF) centered at the two sources, respectively (Fig.~\ref{fig:NGC4435nuclear}a,b). The spectra of src02 can be fitted with an absorbed power law (Fig.~\ref{fig:NGC4435nuclear}d). We do not find any significant time variation or additional foreground absorption toward src02. Therefore, there are just two free parameters: the power law photon index with a best-fit value of $\Gamma=1.61\pm0.13$ (Table~\ref{table:SpecPara}) and the 0.5-7~keV flux with a best-fit value of $\log F_{\rm X}/{\rm (ergs~s^{-1}~cm^{-2})}=-13.47\pm0.04$. The spectra of src01 are more complicated. There is an additional thermal component (fitted with an APEC model) in the spectra of ObsID 2883, but due to the short exposure time of ObsID 8042 and the degrading of the \emph{Chandra}/ACIS in recent years, we do not collect sufficient number of photons to confirm the existence of this component in later observations (Fig.~\ref{fig:NGC4435nuclear}c). The four spectra are still jointly fitted, but only ObsID 2883 has such a thermal component. The best-fit power law photon index is consistent in different observations ($\Gamma=1.51_{-0.75}^{+0.61}$), while the temperature of the thermal component in ObsID 2883 is $kT=0.07_{-0.03}^{+0.09}\rm~keV$ (Table~\ref{table:SpecPara}). We also find significant additional absorption of both components, with $N_{\rm H}=2.8_{-1.7}^{+2.0}\times10^{22}\rm~cm^{-2}$ and $0.96_{-0.50}^{+1.09}\times10^{22}\rm~cm^{-2}$ for the power law and APEC components, respectively (Table~\ref{table:SpecPara}). The significant difference of $N_{\rm H}$ between these two components strongly suggests that a large fraction of the absorption should be intrinsic from the AGN, while some additional absorption from the dusty disk also may be possible.

\subsection{Multi-Wavelength Data} \label{subsec:MultiWavelengthData}

We also obtain some archival multi-wavelength data of NGC~4435/4438, which will be compared to the \emph{Chandra} data in the following sections. In particular, we use broad-band optical images to trace stellar light on different scales, including the high-resolution \emph{HST} image of the nuclear region of NGC~4435 and NGC~4438 (Fig.~\ref{fig:NGC4435nuclear}b, \ref{fig:NGC4438bubblesImg}c), and the low-resolution SDSS image covering both galaxies (Fig.~\ref{fig:NGC4438bubblesImg}a,b). We also use H$\alpha$ narrow-band images to trace the emission from ionized gas. The data presented in this paper are the \emph{HST} image of the nuclear region of NGC~4438 (Fig.~\ref{fig:NGC4438bubblesImg}c; \citealt{Kenney02}), as well as the image taken with the KPNO 4m Mayall telescope (Fig.~\ref{fig:NGC4438bubblesImg}a; \citealt{Kenney08}).

As members of the CHANG-ES (Continuum Halos in Nearby Galaxies: An EVLA Survey) program \citep{Irwin12}, we also show some of our latest images of NGC~4438 observed with the Karl G. Jansky Very Large Array (\emph{VLA}). While most of our results on these \emph{VLA} observations will be presented elsewhere, we herein present our highest resolution images from B-configuration, L-band (centered at 1.5~GHz with 512~MHz bandwidth) and C-configuration, C-band (centered at 6~GHz, 2~GHz bandwidth) data, in Fig.~\ref{fig:NGC4438bubblesImg}b,d. These data were observed on July 29 (B-configuration, L-band) and Feburary 19 (C-configuration, C-band), 2012, respectively, both during the commissioning phase after the upgrade of \emph{VLA} in 2012. General information on the observations and reductions are given in Data Release papers~III \citep{Irwin19} and IV \citep{Walterbos22}. The B-configuration data have a resolution (beam size) of $\sim3^{\prime\prime}$  and a sensitivity of $\sim40\rm~microJy/beam$, while the C-configuration data resolution is $\sim2.7^{\prime\prime}$ and reach a $\sim3.2\rm~microJy/beam$ sensitivity. Both resulting images were weighted with a Briggs robust 0 weighting and have been primary beam corrected. Note that the Virgo region adds complications to radio data handling, because of contamination by strong sources in the field (e.g., M87, which is $\sim1^\circ$ southeast to NGC~4438). As a result, the sensitivity - the rms noise level - for the B-configuration L-band data is about twice that of the expected level for the exposure time (as compared to other CHANG-ES galaxies in less busy fields). 

\section{Spatially Resolved Spectral Analysis of the Nuclear Bubble of NGC~4438} \label{sec:NuclearBubbleN4438}

\begin{figure*}[!th]
\begin{center}
\epsfig{figure=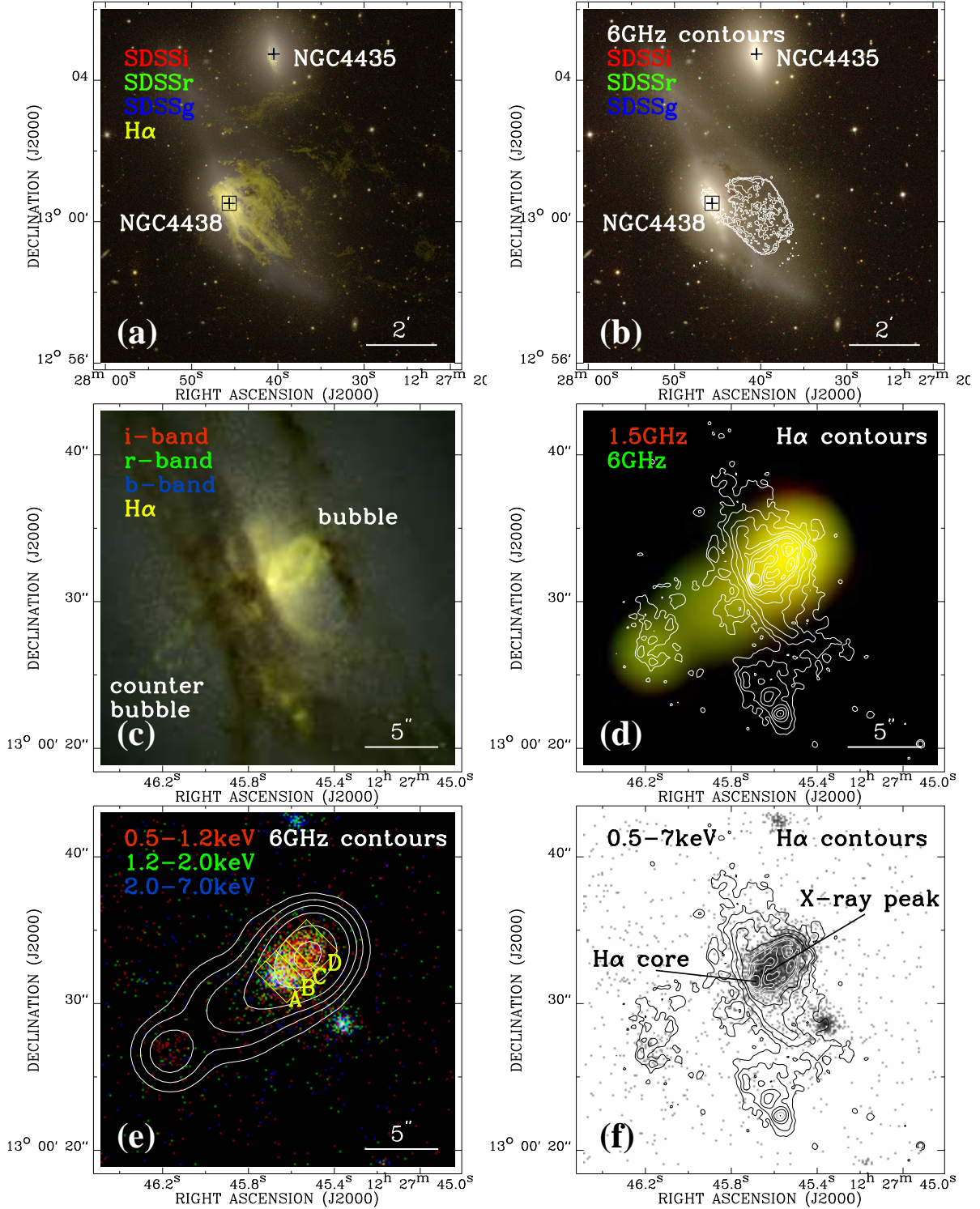,width=0.88\textwidth,angle=0, clip=}
\caption{Multi-wavelength montages of NGC~4438. The images included in these plots are the SDSS g-, i-, r-band; KPNO 4m H$\alpha$ (a); \emph{VLA} B-configuration L-band (1.5~GHz), \emph{VLA} C-configuration C-band (6~GHz); \emph{HST} H$\alpha$ (c), i-, r-, b-bands; \emph{Chandra} broad (0.5-7~keV), soft (0.5-1.2~keV), medium (1.2-2.0~keV), and hard (2.0-7.0~keV) bands, as denoted in each panel. The super-resolution \emph{Chandra} images created with the EDSER algorithm in panel~(e,f) have a pixel size of $0.123^{\prime\prime}$, which is 0.25 times of the original ACIS pixel size. Panel~(a,b) shows the large scale lopsided bubble on the western side, while (c-f) show the nuclear region containing the nuclear bubble with an angular size of $\sim3^{\prime\prime}\times5^{\prime\prime}$, plus a ``counter bubble'' on the opposite side, both detected in H$\alpha$, X-ray, and radio. The small black box in (a, b) shows the $0.4^\prime\times0.4^\prime$ FOV of the following four panels. The four yellow boxes in (e) have the same horizontal extension of $3^{\prime\prime}$, and a vertical extension of $1^{\prime\prime}$, $1^{\prime\prime}$, $1^{\prime\prime}$, and $2^{\prime\prime}$, respectively. These are used to extract the spectra shown in Fig.~\ref{fig:NGC4438bubblesSpec}a-d.
}\label{fig:NGC4438bubblesImg}
\end{center}
\end{figure*}

We present multi-wavelength images of NGC~4438 in Fig.~\ref{fig:NGC4438bubblesImg}. Panels~(a,b) show the large-scale multi-wavelength structures. The most prominent feature is the $\sim2^\prime$-diameter lopsided bubble which has coherent structures in radio continuum (\emph{VLA} C-band contours obtained from the CHANG-ES group; \citealt{Irwin12}), H$\alpha$ \citep{Kenney08}, and X-ray \citep{Machacek04,Li13a}. This large-scale bubble is only detected on the western side, and aligned in a clearly different direction than the small-scale nuclear bubble or the minor axis of the galaxy (shown in Panels~b-d). \citet{Kenney08} discovered some large-scale H$\alpha$ filaments connecting NGC~4438 and M86. The large-scale bubble is roughly aligned along these H$\alpha$ filaments. Therefore, it is likely that the large-scale AGN outflow has been reshaped by the surrounding intracluster medium (ICM). 

As presented in Fig.~\ref{fig:NGC4438bubblesImg}c-f, the $\sim3^{\prime\prime}\times5^{\prime\prime}$ nuclear bubble of NGC~4438 is clearly detected in radio, H$\alpha$, and X-ray. The bubble is also clearly resolved in the \emph{HST} H$\alpha$ image and marginally resolved in the \emph{Chandra} X-ray images created with the EDSER algorithm (\S\ref{subsec:ChandraDataCalibration}). Furthermore, the nuclear bubble of NGC~4438 was also observed by the \emph{VLA} at A-configuration in 1986 (L- and C-band) and 1998 (X-band) before the 2012 upgrade of the observatory \citep{Hummel91,Hota07}. The nuclear bubble is clearly resolved in C-band (4.86~GHz) and X-band (8.46~GHz) into two shells, which match the structures revealed in H$\alpha$ and X-ray (\citealt{Hota07}; Fig.~\ref{fig:NGC4438bubblesImg}c,e,f). A weaker ``counter bubble'' is also clearly detected in the three bands shown in Fig.~\ref{fig:NGC4438bubblesImg}c,e,f. There is a weak point-like source coinciding with the galactic nucleus revealed in the high-resolution \emph{VLA} A-configuration images presented in \citet{Hota07}. Thanks to the new \emph{Chandra} data, we now have more X-ray photons than \citet{Machacek04}, especially in the hard X-ray band. The peak of the X-ray emission is resolved to be clearly offset from the compact core in H$\alpha$ (Fig.~\ref{fig:NGC4438bubblesImg}f). This compact core also likely coincides with the radio core resolved in the \emph{VLA} A-configuration images \citep{Hota07}. The X-ray emission also shows some visible structures, instead of being purely point-like. Therefore, we regard it as extended emission from the bubble, rather than from the AGN. 

\begin{figure*}[!th]
\begin{center}
\epsfig{figure=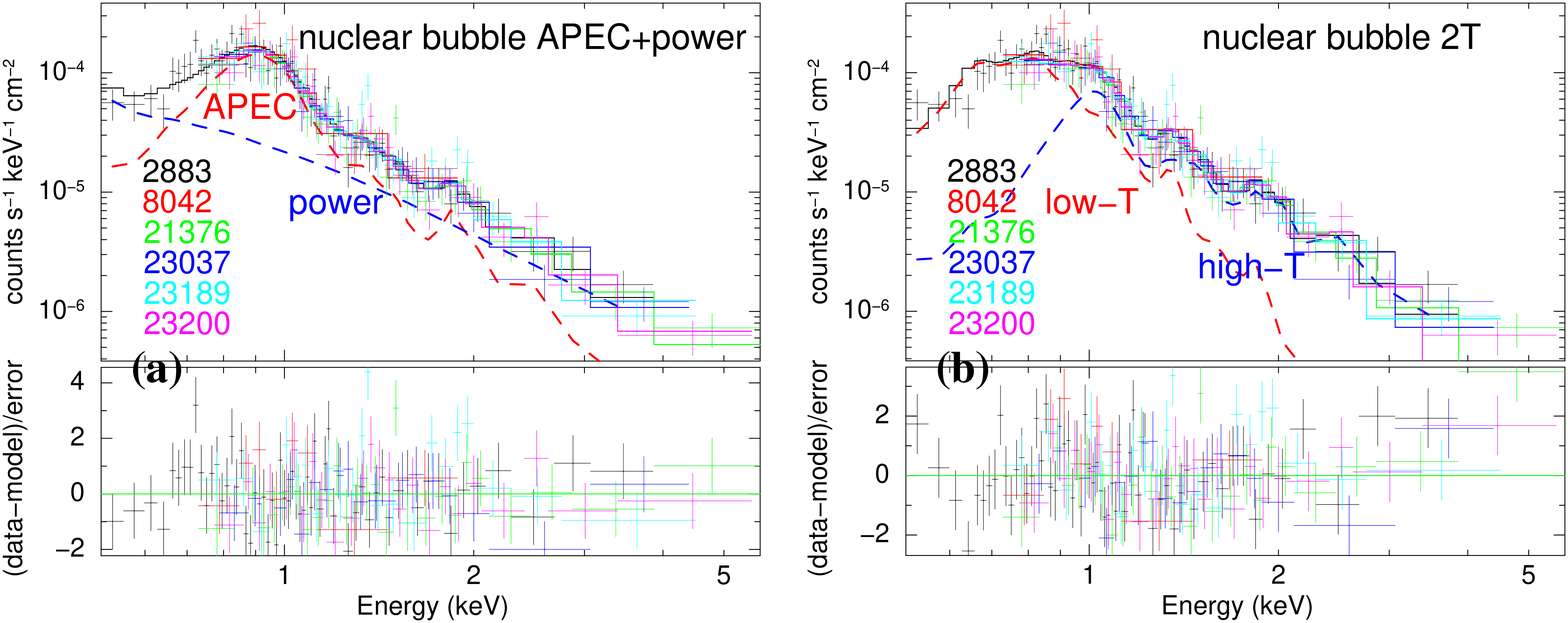,width=1.0\textwidth,angle=0, clip=}
\caption{\emph{Chandra} spectra extracted from a $\sim3^{\prime\prime}\times5^{\prime\prime}$ elliptical region roughly covering the entire bubble of NGC~4438 (Fig.~\ref{fig:NGC4438bubblesImg}e). In the two panels, the spectra are fitted with two different models: ``APEC+power law'' (a) and ``APEC+APEC'' (or 2T model; b), with different model components plotted in different colors. The best-fit model parameters are listed in Table~\ref{table:SpecPara}. The observed spectra were corrected for the effective area, using the XSpec plot setting ``setplot area''. This setting is helpful to compare observations taken in different epoch (the first two observations are clearly more sensitive than the latest observations taken after the significant degrading of ACIS), and to show the decomposition of different model components.}\label{fig:NGC4438bubbleModel}
\end{center}
\end{figure*}

\begin{figure*}[!th]
\begin{center}
\epsfig{figure=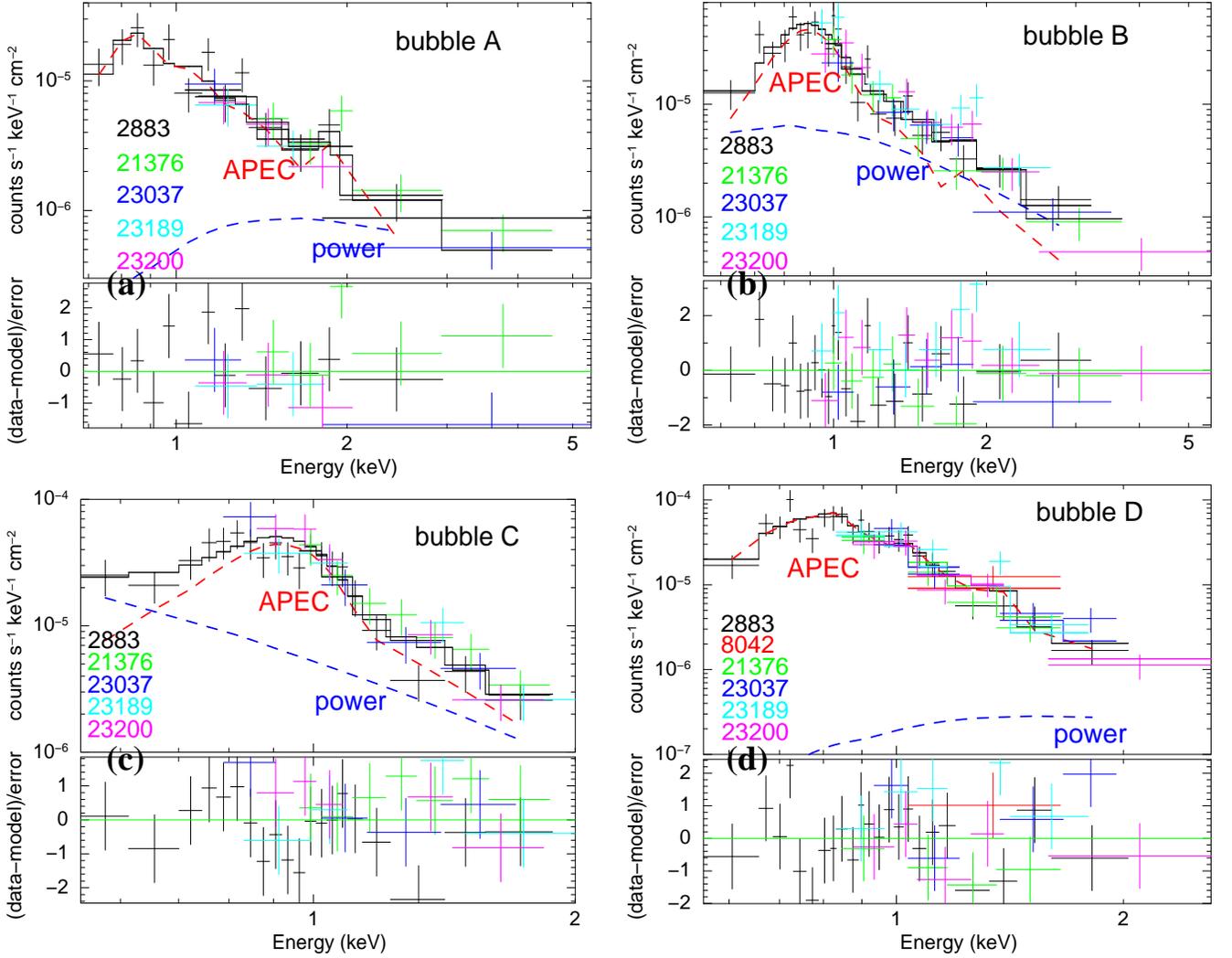,width=1.0\textwidth,angle=0, clip=}
\caption{\emph{Chandra} spectra of the nuclear bubble of NGC~4438 extracted from the $3^{\prime\prime}$-wide box regions (Fig.~\ref{fig:NGC4438bubblesImg}e) with an increasing vertical distance from the galactic plane. The observations used in each panel are denoted on the bottom left (Table~\ref{table:ChandraObs}). In some regions, there are too few counts from ObsID 8042, so this observation is not included in the spectral analysis. The red and blue dashed curves are the thermal plasma (APEC) and power law components, respectively.
}\label{fig:NGC4438bubblesSpec}
\end{center}
\end{figure*}

We conduct spatially resolved spectroscopy analysis of the nuclear bubble of NGC~4438 with the \emph{Chandra} data and present the results in Figs.~\ref{fig:NGC4438bubbleModel} and \ref{fig:NGC4438bubblesSpec}. Some of the key best-fit model parameters are summarized in Table~\ref{table:SpecPara}. We first analyze the \emph{Chandra} spectra extracted from the entire nuclear bubble (Fig.~\ref{fig:NGC4438bubbleModel}), which is extracted from a $\sim3^{\prime\prime}\times5^{\prime\prime}$ elliptical region. The sky background is extracted from a source-free region far away from the galaxy on the ACIS-S3 chip, and is identical for all the observations. The spectra extracted from each observations have been adaptively regrouped to have a minimum number of 10 photons in each energy bin (the same in the following analysis). Since the spectra show a clear featureless high-energy ``tail'' which cannot be well fitted with a single-temperature thermal plasma component, we fit the spectra extracted from different observations jointly in XSpec with two different models: an ``APEC+power law'' model (Fig.~\ref{fig:NGC4438bubbleModel}a) and a ``2T'' model which is comprised of two APEC components with different temperatures (Fig.~\ref{fig:NGC4438bubbleModel}b). In any of these two models, both model components are subjected to foreground extinction described with the ``tbabs'' model. The APEC component is used to describe the thermal plasma contribution, while the power law component could be used to describe any non-thermal contributions (e.g., synchrotron emission). 

The two models have the same degree of freedom, but the ``APEC+power law'' model provides a better fit with a $\chi^2/d.o.f=163.12/156$, compared to 193.77/156 of the ``2T'' model (Table~\ref{table:SpecPara}). The ``2T'' model also shows a significant residual at $\gtrsim3\rm~keV$ and also poorer fit at lower energies. Therefore, we adopt the ``APEC+power law'' model in the following analysis and discussions, which is the simplest description of the X-ray spectra also adopted in previous works (e.g., \citealt{Machacek04}). However, we emphasize that both models tested here are over simplified, so we cannot firmly rule out any of them with the current data. A thermal plasma model with a broader temperature distribution but no non-thermal component is still possible. In particular, a high-temperature thermal plasma could present, especially when there is a strong magnetic confinement effect (e.g., \citealt{Wang21}).   

\begin{figure*}[!th]
\begin{center}
\epsfig{figure=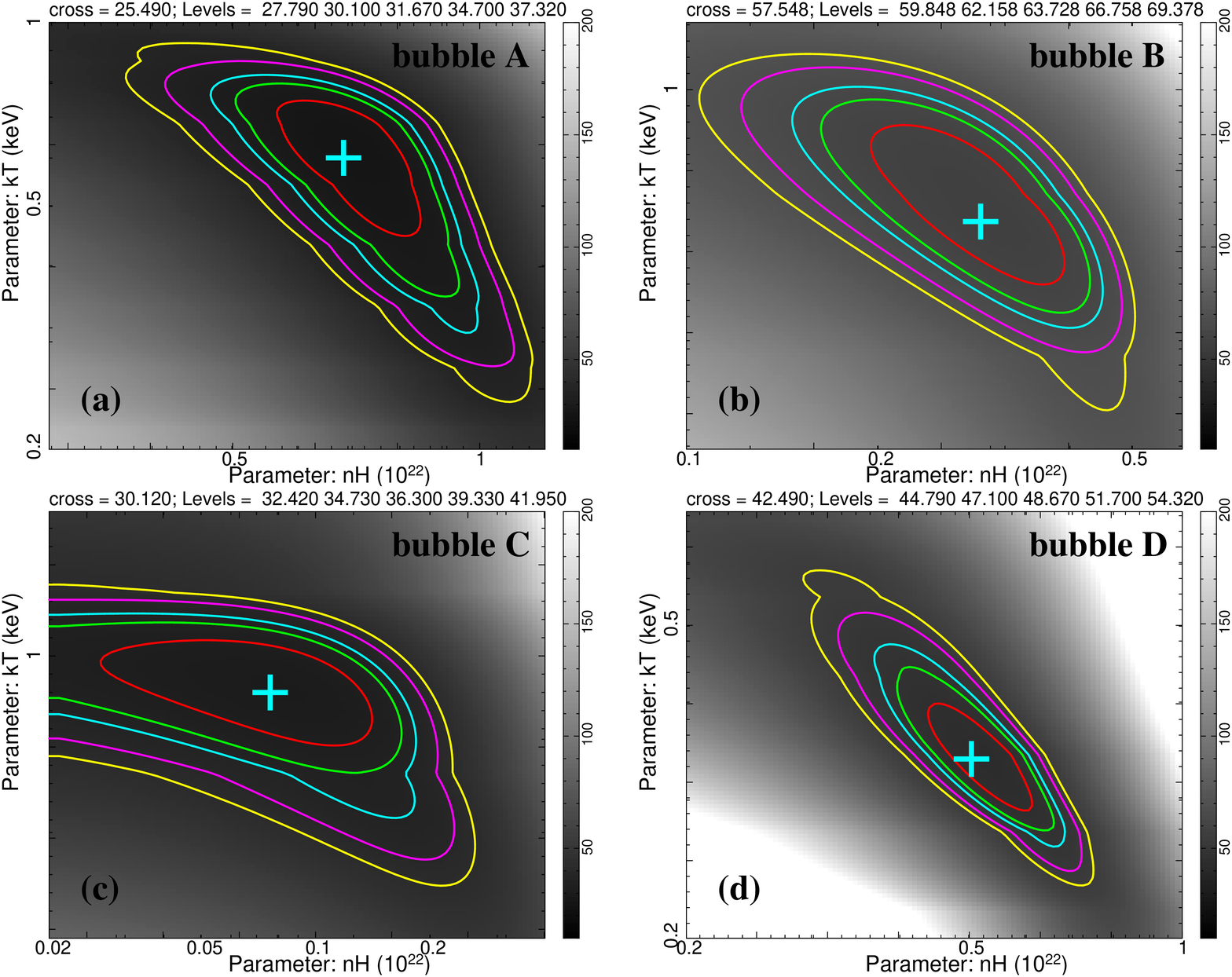,width=1.0\textwidth,angle=0, clip=}
\caption{The $N_{\rm H}$ (plotted as nH) vs $kT$ parameter space of the spectra shown in Fig.~\ref{fig:NGC4438bubblesSpec}. The contours are at levels of 1-$\sigma$, 90\%, 2-$\sigma$, 99\%, and 3-$\sigma$. The two parameters are in general anti-correlated with each other, so not completely independent. The respective axis-projections of the innermost contour (red) correlate with the errors listed in Table~\ref{table:SpecPara} and plotted in Fig.~\ref{fig:NGC4438bubblesProfile}.
}\label{fig:NGC4438bubbleParaSpace}
\end{center}
\end{figure*}

We further study the spatial variation of the physical parameters of the nuclear bubble obtained from X-ray spectral analysis. We extract the \emph{Chandra} spectra from the four box regions aligned roughly along the minor axis of the galaxy, covering the nuclear bubble (Fig.~\ref{fig:NGC4438bubblesImg}e; labeled as ``bubble A, B, C, D'' with increasing distance from the nucleus). The horizontal size of all these box regions is $3^{\prime\prime}$ ($1^{\prime\prime}\approx70\rm~pc$), while the vertical size is $1^{\prime\prime}$ for the first three regions and $2^{\prime\prime}$ for the last one. The larger size of the spectral extraction region for ``bubble D'' is to ensure a sufficient number of photons for spectral analysis. We follow the same criteria in binning the spectra extracted from different regions and different epochs. Individual spectra were kept in the fitting if they have at least one usable bin in the energy range of interest. For ``bubble A-C'', we do not have a sufficient number of photons from ObsID 8042, so only five observations are used in the spectral analysis (Fig.~\ref{fig:NGC4438bubblesSpec}a-c). All the spectra are fitted with the same model we adopted for the entire bubble (Fig.~\ref{fig:NGC4438bubbleModel}a). Some best-fit parameters are summarized in Table~\ref{table:SpecPara}. Some regions (bubble~A-C) show a significant non-thermal tail which cannot be well fitted with a single-temperature thermal plasma model. Such a non-thermal tail is quite weak in the outermost region (bubble~D).

Some of the model parameters may be degenerated in the analysis of the X-ray spectra with low counting statistics. We show the confidence range of the temperature of the APEC component ($kT$) and the absorption column density ($N_{\rm H}$) of different regions in Fig.~\ref{fig:NGC4438bubbleParaSpace}. These two parameters appear to have a fairly good anti-correlation, so they are not completely independent in the fitting of low-quality spectra (Fig.~\ref{fig:NGC4438bubblesSpec}). The actual uncertainty (including the systematic uncertainties caused by these degeneracies or the spectral decomposition) of the best-fit parameters could thus be significantly larger than shown in Table~\ref{table:SpecPara}. However, we indeed see a systematic shift of the peak of the X-ray spectra among different regions (Fig.~\ref{fig:NGC4438bubblesSpec}), which indicates the possible spatial variations of the hot gas and non-thermal emission properties. Such spatial variations will be discussed in \S\ref{sec:Discussions}, under the assumptions of spectral analysis as discussed above.

\begin{figure*}[!th]
\begin{center}
\epsfig{figure=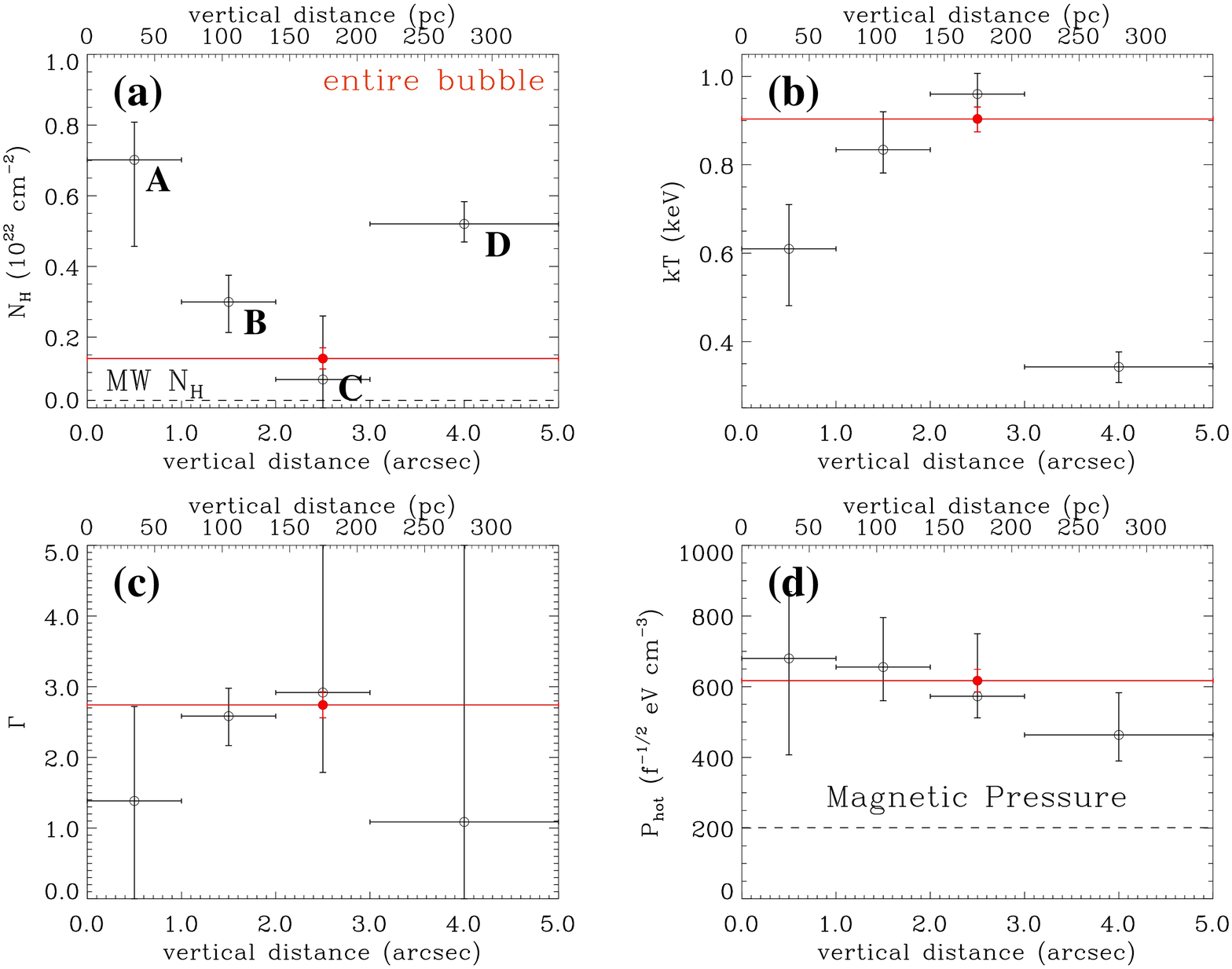,width=1.0\textwidth,angle=0, clip=}
\caption{Vertical variation of a few X-ray parameters obtained by analyzing the \emph{Chandra} spectra in Fig.~\ref{fig:NGC4438bubblesSpec}: foreground absorption column density $N_{\rm H}$ (a), hot gas temperature $kT$ (b), power law photon index $\Gamma$ (c), and hot gas thermal pressure $P_{\rm hot}$ (d). The red data point with large horizontal error bar is the best-fit value of the entire bubble as shown in Fig.~\ref{fig:NGC4438bubbleModel}a. We mark in (a) the labels ``A, B, C, D'' of different small regions as plotted in Fig.~\ref{fig:NGC4438bubblesImg}e. We also plot in (a) the level of the MW foreground absorption toward NGC~4438, and in (d) the magnetic pressure assuming $B=90\rm~\mu G$ \citep{Hota07}, as dashed lines for comparison.
}\label{fig:NGC4438bubblesProfile}
\end{center}
\end{figure*}

\section{Discussions} \label{sec:Discussions}

Based on our knowledge, NGC~4438 is one of the only two cases where we firmly detect diffuse non-thermal X-ray emission associated with a nuclear bubble in an external galaxy (see detection of the diffuse hard X-ray emission from other galaxies in \citealt{Lacki13}). In another case of NGC~3079, because the cosmic microwave background (CMB) is not strong enough compared to the magnetic field [so the Inverse Compton (IC) emission is relatively weak], such a non-thermal component has been interpreted as the synchrotron emission from $\sim(10-100)\rm~TeV$ CR leptons \citep{Li19}. Typically, only when the magnetic field strength $B\lesssim3\rm~\mu G$ could the IC emission be more important (e.g., \citealt{Reynolds98}). \citet{Hota07} estimated the magnetic field strength based on their \emph{VLA} A-configuration observations and an energy equipartition assumption between CR and magnetic field. They obtained a magnetic field strength of the nuclear bubble of NGC~4438 of $B\sim90\rm~\mu G$, which is far larger than the upper limit where IC emission can be more important. We therefore assume the entire non-thermal power law component is produced by the synchrotron emission of TeV CR leptons. Although the strength of this component is highly uncertain and depends on how we decompose the X-ray spectra, we mainly adopt the best-fit parameters listed in Table~\ref{table:SpecPara} in the following discussions. 

We present the vertical variation of $N_{\rm H}$, $kT$, and the power law photon index $\Gamma$ of the nuclear bubble in Fig.~\ref{fig:NGC4438bubblesProfile}a-c. Due to the negligible contribution of the power law component in bubble D (Fig.~\ref{fig:NGC4438bubblesSpec}d), the error bar of the last data point in Fig.~\ref{fig:NGC4438bubblesProfile}c is very large. Also because of the degeneracy of $N_{\rm H}$ and $kT$ (Fig.~\ref{fig:NGC4438bubbleParaSpace}), the errors could be larger than presented in the figure when the counting statistic is poor (especially for bubble D). Nevertheless, we still see some significant trends from these vertical profiles. $N_{\rm H}$ is decreasing, while $kT$ and $\Gamma$ are both increasing with the vertical distance in the first three data points. The significantly decreasing $N_{\rm H}$ is consistent with the nearly edge-on orientation of the galaxy, and almost reach the Milky Way (MW) foreground value at $h\sim3^{\prime\prime}\sim200\rm~pc$ from the nucleus. The high $N_{\rm H}$ of bubble~D is somewhat unexpected, which could be caused by some gas clumps projected at front, or simply a result of the large measurement uncertainty owing to the $N_{\rm H}$-$kT$ degeneracy discussed above (the $N_{\rm H}$-$kT$ degeneracy is stronger in bubble~D than in  other regions; Fig.~\ref{fig:NGC4438bubbleParaSpace}). Furthermore, $kT$ of bubble~D is also significantly lower than the other regions (Fig.~\ref{fig:NGC4438bubblesProfile}b), consistent with the scenario that the higher $N_{\rm H}$ of bubble~D could at least be partially caused by the $N_{\rm H}$-$kT$ degeneracy. 

The significant increase of the gas temperature at $h\lesssim200\rm~pc$ from the nucleus strongly suggests that the hot gas is heated when the bubble is blowing out. On the other hand, there may be a coherent increase of $\Gamma$ together with the increase of the gas temperature at $h\lesssim200\rm~pc$ (Fig.~\ref{fig:NGC4438bubblesProfile}c), although the uncertainty in spectral decomposition results in the large measurement error and the results are not inconsistent with a constant non-thermal spectral slope. The vertical increase of $\Gamma$, if real, indicates the CRs are losing energy. This apparently indicates that the CRs may be transferring energy to the hot gas in some ways (e.g., by hadronic or Coulomb interactions, or the damping of Alfv$\rm\acute{e}$n waves excited by streaming CRs). However, these processes are often efficient only on larger scales in the hot gas (e.g., a few tens to a few hundreds kpc; \citealt{Uhlig12,Ruszkowski17}), and we do not have any strong evidence that the two processes are physically coupled to each other. The vertical variations of $kT$ and $\Gamma$ may simply be interpreted as the simultaneous shock heating of the ambient gas and the synchrotron radiative cooling of the CR leptons.

We estimate the synchrotron cooling timescale of the CR leptons responsible for the non-thermal X-ray emission. The synchrotron cooling timescale $t_{\rm syn}$ in a magnetic field $B$ at the peak synchrotron emission frequency $\nu_{\rm m}$ can be described as: 
\begin{equation}\label{Equ:tsyn}
\begin{aligned}
t_{\rm syn}/{\rm s}=~&8.7\times10^{11}\times(B/{\rm Gauss})^{-3/2}\\ 
&\times(\nu_{\rm m}/{\rm Hz})^{-1/2}\times(\sin\alpha)^{-3/2}, 
\end{aligned}
\end{equation}
where $\alpha$ is the angle between the moving direction of the CR lepton and the magnetic field (Eq. 4.54 of \citealt{You98}; also see \citealt{Krause18}). Adopting $B\sim90\rm~\mu G$ estimated from \citet{Hota07}, the peak frequency of a $4\rm~keV$ photon ($\nu_{\rm m}\sim10^{18}\rm~Hz$), and $\alpha=90^\circ$, we obtain a typical synchrotron cooling timescale of the CR leptons responsible for the observed diffuse hard X-ray emission $t_{\rm syn}\sim34\rm~yr$. Considering the large uncertainty of the estimated magnetic field (e.g., for $B\sim20-200\rm~\mu G$, we obtain $t_{\rm syn}\sim10-300\rm~yr$), the typical value of the synchrotron cooling timescale should be a few tens to a few hundreds of years. The outflowing velocity of CRs is largely determined by the propagation mechanisms (diffusion or advection dominated), but is typically much lower than the speed of light for particles at TeV energy (e.g., thermally driven outflow typically have a velocity $\lesssim10^3\rm~km~s^{-1}$). Therefore, a TeV CR particle cannot travel a few hundred pc before losing a significant fraction of its energy. This excludes the possibility that the CRs are directly accelerated by the AGN. 

The angular resolution of the \emph{Chandra} image is insufficient to confirm where the CRs are accelerated around the base of the bubble. The limited number of X-ray photons also makes the decomposition of the thermal and non-thermal components quite uncertain. Although the non-thermal component seems significant in the spectra from bubble A to C (Fig.~\ref{fig:NGC4438bubblesSpec}), the vertical variations of its relative contribution and spectral slope are quite artificial. Therefore, we cannot rule out the possibility that the CRs are accelerated by some mechanisms closely related to the AGN (e.g., by the direct interaction between the AGN outflow or jet and the surrounding medium, instead of the AGN itself). They then transport outward, sometimes may get re-accelerated, and produce the non-thermal X-ray emissions. An alternative scenario is that the CRs are accelerated in situ at the shell of the bubble, as suggested by \citet{Li19} in NGC~3079. However, it is clear that the outer shell or the ``cap'' of the bubble (bubble D; Fig.~\ref{fig:NGC4438bubblesSpec}d) is dominated by thermal emissions, without significant signatures of CR re-acceleration to such high energies. Our observational constraint on the extension and spectral slope of the non-thermal hard X-ray emission around the AGN blowout bubble of NGC~4438 may help to constrain the CR acceleration, propagation, and cooling in some leptonic AGN bubble models (e.g., \citealt{Yang17}).  

We present the thermal pressure profile in Fig.~\ref{fig:NGC4438bubblesProfile}d and perform a rough examination of the pressure balance between the magnetic field and the hot gas. However, we caution that both the estimated magnetic field strength in \citet{Hota07} and the hot gas parameters from this paper (Table~\ref{table:SpecPara}) have large uncertainties. In particular, it is often difficult to reach a balance between the CRs and magnetic field on a scale of just $\sim300\rm~pc$ (e.g., \citealt{Seta19}), so the energy equipartition assumption may not be correct for the nuclear bubble. 

We first calculate the electron density of hot gas ($n_{\rm e}$) from the fitted volume emission measure of the APEC component ($EM_{\rm hot}$), assuming a volume filling factor $f$ and a box region with the length along the line of sight equal to $3^{\prime\prime}$ (assuming ellipsoid region for the entire bubble). We then calculate the thermal pressure of the hot gas ($P_{\rm hot}$) using $n_{\rm e}$ and the fitted hot gas temperature $kT$. The derived thermal pressure of the hot gas within the nuclear bubble is typically in the range of $\sim(400-800)\rm~eV~cm^{-3}$, assuming $f\sim1$ (Table~\ref{table:SpecPara}; Fig.~\ref{fig:NGC4438bubblesProfile}d). On the other hand, adopting the magnetic field strength of $B\sim90\rm~\mu G$ from \citet{Hota07}, the magnetic pressure of the nuclear bubble is $\approx200\rm~eV~cm^{-3}$ ($\approx3\times10^{-10}\rm~dyn~cm^{-2}$), about three times lower than the typical value of the thermal pressure. 

The spectral fitting of $EM_{\rm hot}$ and the calculation of $n_{\rm e}$ is highly affected by $kT$ (higher $kT$ results in lower $EM_{\rm hot}$ and $n_{\rm e}$), so the $kT$-$N_{\rm H}$ degeneracy as discussed above. Nevertheless, the thermal pressure is in general less affected, because $P_{\rm hot}\propto n_{\rm e}kT$, while $n_{\rm e}$ and $kT$ are anti-correlated with each other in the spectral fitting. Furthermore, a lower filling factor than the assumed value of $f=1$ will only result in an even higher $P_{\rm hot}$. Therefore, we do not expect a significantly lower thermal pressure of the nuclear bubble than we estimated above. The adopted magnetic field strength is an average value of the bubble, and is based on many assumptions such as the energy equipartition between CRs and magnetic field, as well as the ratio of heavy particle to electron number density in the 
GeV energy range $k$ (\citealt{Hota07} obtained a magnetic field strength of $B=68.6\rm~\mu G$ assuming $k=40$, while $B=88.8\rm~\mu G$ if $k=100$; we assume $B=90\rm~\mu G$ for a moderate shock strength in this paper), so it is not impossible that the real magnetic field is stronger. Better radio observations, especially polarization and rotation measure synthesis (e.g., \citealt{MoraPartiarroyo19a,MoraPartiarroyo19b,Krause20,Stein20}), are needed to better estimate the magnetic field strength of the nuclear bubble, in order to examine the pressure balance between the hot gas and the magnetic field.

Detection of the extended non-thermal hard X-ray emissions associated with galactic nuclear superbubbles is critical in understanding how the high energy CRs are originally accelerated. Compared to another case of hard X-ray emitting superbubble detected in NGC~3079 \citep{Li19}, the hard X-ray emission in the nuclear bubble of NGC~4438 is less extended ($\sim100\rm~pc$ vs $\sim1\rm~kpc$). This causes some confusions in attributing the hard X-ray emission to the bubble instead of the AGN. NGC~4438 does not have strong star formation activities, as indicated by its low measured SFR \citep{Vargas19} and unusually high radio-to-IR flux ratio \citep{Li16}. On the other hand, NGC~3079 is very active in both AGN and star formation. It is still not clear if the more extended hard X-ray emission in NGC~3079 is caused by starburst feedback or not. Another difference between these two galaxies is the environment, with NGC~3079 being a field galaxy with a few small companions, but NGC~4438 locates in the Virgo cluster and highly disrupted by the companion NGC~4435 and M86 (e.g., \citealt{Li13a}). The dense ICM could play a critical role in reshaping the multi-scale bubbles around NGC~4438 (e.g., Fig.~\ref{fig:NGC4438bubblesImg}a,b). In particular, the strong nuclear bubble in the NE direction is roughly aligned with the large-scale H$\alpha$ filaments connecting NGC~4438 and M86 \citep{Kenney08}, while the much fainter counter bubble is likely penetrating into a lower density ICM. It is possible but not confirmed that the dense environment plays a role to help thermalize the energy released by the AGN outflow, so produce less hard X-ray emissions than in NGC~3079. Non-thermal emissions, especially in radio band, have been detected around many galaxies with or without a nuclear bubble (e.g., \citealt{Irwin12,Wiegert15,Krause18,Heald22}). However, the typical energy of CR leptons responsible for these radio emissions is $\sim\rm GeV$, which is much lower than the TeV CRs responsible for the hard X-ray emission. Since the low-energy CRs have a much longer lifetime, they usually cannot be used to trace the re-acceleration of CRs on the superbubble scale. Therefore, in many leptonic CR models (e.g., \citealt{Yang17}), such as for the ``Fermi bubble'' of the MW, the extended broad-band non-thermal emissions could be produced by the CRs accelerated in the nuclear region, which then propagate to the bubble shell.

\section{Summary and Conclusions} \label{Sec:SummaryConclusion}

In this paper, we present the latest \emph{Chandra} observations of the Virgo cluster galaxy NGC~4438. We study some interesting point-like sources in the nuclear region of NGC~4438 and its companion galaxy NGC~4435. In particular, we discover a transient X-ray source only $\sim5^{\prime\prime}$ from the nucleus of NGC~4438. The source was not detected in 2002 and 2008, but became quite X-ray bright in March 2020, with an average 0.5-7~keV luminosity of $\sim10^{39}\rm~ergs~s^{-1}$, assuming the same distance as NGC~4438. We also detect two X-ray bright sources close to the nucleus of NGC~4435, and identify one of them as its AGN. The X-ray spectra of the AGN show a clear thermal component in addition to the power law component. Both components are subjected to strong foreground extinction, which are much larger than the value of another nearby source. However, the extinction values of the thermal and power law components of the AGN are significantly different.

NGC~4438 hosts a $\sim2^\prime$ ($\sim4\rm~kpc$)-diameter lopsided bubble to the west of the warped disk, as well as a nuclear $\sim3^{\prime\prime}\times5^{\prime\prime}$ ($\sim200{\rm~pc}\times350\rm~pc$) bubble with a much fainter counterpart on the opposite side. All these multi-scale features are detected to have coherent structures in radio continuum, H$\alpha$, and soft X-rays. In particular, the nuclear bubble is also detected in hard X-rays at $>2\rm~keV$, which makes it one of the only two external galactic nuclear bubbles with associated extended hard X-ray emissions.

X-ray spectral analysis of the nuclear bubble indicates significant non-thermal tails at $\gtrsim2\rm~keV$, which is most naturally interpreted as synchrotron emission of high-energy [$\sim(10-100)\rm~TeV$] CR leptons. Our spatially resolved spectroscopic analysis further indicates that the hot gas temperature is increasing while the overall contribution from the non-thermal X-ray emission is decreasing with increasing vertical distances from the galactic plane. Within a vertical height of $h\lesssim200\rm~pc$, the shape of the non-thermal X-ray emission may also steepen. If this trend can be confirmed, it may indicate a strong synchrotron cooling of CR leptons. Adopting the magnetic field strength estimated from the high-resolution radio data, we found that the synchrotron cooling timescale of the CR leptons responsible for the non-thermal hard X-ray emission is only a few tens to a few hundreds of years. The current data are not good enough to distinguish if the CRs are accelerated in situ at the bubble shell or accelerated by the AGN then transported to the place where they radiate synchrotron emission. The AGN acceleration scenario seems to be consistent with the apparent vertical steepening of the non-thermal X-ray spectra (if confirmed) and the lack of hard X-ray emission from the ``cap'' of the bubble.

We examine the pressure balance between the hot gas and the magnetic field, based on both the X-ray data and the estimate of magnetic field strength in the archive. The average hot gas thermal pressure is about five times the magnetic pressure, but both quantities could have large uncertainties, so we cannot rule out the possibility that they are in pressure balance. Higher resolution radio observations with broad frequency coverage and polarization are needed to directly estimate the magnetic field strength and structure associated with the bubble.

\acknowledgments
The authors acknowledge Prof. Jeffrey Kenney for providing the reduced H$\alpha$ images from \emph{HST} and the KPNO 4m Mayall telescope. The authors also acknowledge the financial support from NASA and the \emph{Chandra} X-ray Center through the grant GO9-20074X. TW acknowledges financial support from The Coordination of the Participation in SKA-SPAIN, financed by the Ministry of Science and Innovation (MCIN), from the State Agency for Research of the Spanish Ministry of Science, Innovation and Universities through the ``Centre of Excellence Severo Ochoa" award to the Instituto de Astrof$\rm\acute{i}$sica de Andaluc$\rm\acute{i}$a (SEV-2017-0709). 

Data Availability

The X-ray data underlying this article are available in the \emph{Chandra} data archive, at \url{https://cxc.harvard.edu/cda/}. The radio data underlying this article are available in the CHANG-ES website, at \url{https://www.queensu.ca/changes/}, and will be shared by reasonable request to the corresponding author.

\end{document}